# Exploring AI Futures Through Role Play

Shahar Avin
Centre for the Study of Existential Risk
University of Cambridge
Cambridge, UK
sa478@cam.ac.uk

Ross Gruetzemacher
Raymond J. Harbert College of Business
Auburn University
Auburn, Alabama, USA
rossg@auburn.edu

James Fox
Department of Engineering
University of Oxford
Oxford, UK
james.fox@keble.ox.ac.uk

## ABSTRACT

We present an innovative methodology for studying and teaching the impacts of AI through a role-play game. The game serves two primary purposes: 1) training AI developers and AI policy professionals to reflect on and prepare for future social and ethical challenges related to AI and 2) exploring possible futures involving AI technology development, deployment, social impacts, and governance. While the game currently focuses on the inter-relations between short-, mid- and long-term impacts of AI, it has potential to be adapted for a broad range of scenarios, exploring in greater depths issues of AI policy research and affording training within organizations. The game presented here has undergone two years of development and has been tested through over 30 events involving between 3 and 70 participants. The game is under active development, but preliminary findings suggest that role-play is a promising methodology for both exploring AI futures and training individuals and organizations in thinking about, and reflecting on, the impacts of AI and strategic mistakes that can be avoided today.

## CCS CONCEPTS

• **Computing methodologies~Artificial intelligence**
• **Social and professional topics~Informal education**
• Social and professional topics~Computing / technology policy
• Social and professional topics~Governmental regulations

## 1 Introduction

Artificial intelligence (AI) has an enormous potential to have a radical effect on our future, and we all have a vested interest in trying to ensure that AI is beneficial for humanity. However, there is currently no consensus in the AI research community on what the long-term impacts of AI, potentially including general artificial intelligence, will be, or the length of time over which such changes will take place. Moreover, there are many near- and mid-term concerns that have the potential for transformative societal change [5, 14]. The vast uncertainty around AI futures, especially ones that involve transformative societal changes, has precipitated the creation of an AI strategy research community that seeks to ensure the development of AI for the social good [8]. In collaboration with researchers in this community, we have developed an AI futures role-play game, which we present here, to complement the research being done and to disseminate key questions and findings to individuals and organizations engaged with, or seeking to engage with, AI strategy and policy.

A role-play game involves a collection of people assuming roles to enact a situation in a realistic manner so as to predict a possible future if certain strategies are employed. Role-plays have a track record of high forecasting, strategic and educational utility in a wide variety of fields, including business [15, 21], medicine [18], politics [16] and education [6, 21]. In the domain of AI strategy, we believe that role-plays can illuminate plausible futures, identify critical decision points, educate, raise awareness, highlight gaps in research knowledge and promote shared visions for beneficial AI.

In this paper we first review the wargaming literature to show where role-plays have been successfully used in other domains, before next looking at why role-plays offer particular benefits within the AI strategy space. We then describe the current structure of the game (still under active development), and we highlight some of the lessons learned from the 30 game events to date. We conclude with a discussion of the benefits and shortcomings of this AI role-play game, our plans for further development and dissemination, and a call for others to try role-play as a methodology for exploring AI futures.

## 2 Literature Review

"Serious" role-plays are most well known in their military form, wargames, where they have been used, since their Prussian development in the late 18th century, to teach military strategy in a safe but dynamic manner. Wilkinson [24] championed the introduction of wargames into the British army in the late 1800's and claimed that "probably no form of military study is more useful if properly conducted". Original wargames had two militaries (teams of players) fighting over some well-defined territory (the board). More recently, emerging technologies and distributed warfare necessitated wargames to shift focus from territory and troop movement to also include the diplomacy of international relations [22, 23], nuclear conflict escalation [20] and cyber operations [1].

The potential of role-plays to provide exploration and training in uncertain and high stakes situations has been recognized in other fields. In business role-plays, for example, the conflict arises between a company and its competitors vying for market share and profit. Hamel and Prahalad [15] explain the popularity of role-plays in business as due to their ability to develop foresight, an essential skill for competitive survival: "an organization which does not develop a picture of the future, which does not develop foresight, is most likely not involved in the future." In education [6, 21], medicine [18] and politics [16], role-plays are used for formulating strategy, developing foresight, crisis response preparation, and for training and recruitment exercises. Nevertheless, despite their proven utility in a variety of different fields, role-plays have so far not been used in AI futures studies [3].

### 2.1 When are role plays suitable?

Parson [19] claims that role-plays are most appropriate for situations involving high enough stakes to merit substantial investment in knowledge about decision-making, but whose complexity imposes limits on the usefulness of standard decision-making procedures. Parson [19] gives three suitability conditions for role-plays:

1. Key outcomes depend on the interacting decisions of multiple agents.
2. Decision choice sets are ambiguous or poorly known.
3. Large numbers of complex organizations may be required to work together.

The existence of multiple interacting complex agents compounded with a huge set of possible decisions creates an enormous potential scenario space. It is too difficult for one expert alone to think beyond linear views of reality, but role-plays allow non-linear effects to be explored. For instance, one can look for convergent outcomes arising from different strategies or identify the decisions pivotal in shaping the scenario's end state. Armstrong [2] compares and contrasts different forecasting techniques and adds that a role-play needs:

4. Large changes expected.
5. Conflict between the decision makers.

If only small changes are expected, forecasters can extrapolate from current or historical examples. However, large technological step-changes – such as the discovery of electricity or the invention of the world wide web - entirely transform history's trajectory and are much harder to predict. This observation has been captured in Roy Amara's law [17]: "we tend to overestimate the effect of a technology in the short run and underestimate the effect in the long run."

Finally, conflict generates oppositional friction which catalyzes the generation of creative ideas. Without the competitive nature of an adversarial interaction, it is hard to generate the surprising edge-case scenarios which are exactly the ones hoped to be reached by role-plays.

### 2.2 Do They Work?

There is empirical evidence substantiating role-playing's greater forecasting accuracy over unaided judgements in certain situations. Green, Graefe and Armstrong [11] showed that both experts and novices were terrible at forecasting conflict decisions if they were simply asked to "put yourself in the other person's shoes". This is because, regardless of expertise, it is too difficult to individually think through the interactions of competing characters with divergent goals in complex situations. However, if the subjects adopted the roles of the protagonists in the conflicts and interacted with each other, their group decisions predicted the actual decisions with much greater accuracy – the greater realism afforded by role-playing enabled superior forecasts to be made.

Role-plays can also mitigate against myopia by helping overcome various different psychological biases. Cyert, March and Starbuck [7] found that subjects presented with the same data produced significantly different forecasts depending on which role they were told to act out and Babcock et al. [4] evidenced the tendency of role-players to make biased judgements and interpret the briefing material differently according to the role they were assigned. This explains why role-playing benefits from larger groups and greater diversity. Duke and Geurts [10] found that role-playing activities in a political setting can mitigate the danger of 'groupthink', where only readily politically feasible or easy-to-implement strategies are discussed.

It is not guaranteed that every role-play will provide novel insights and any insights that do arise might not be relevant or applicable. However, even end-game states that have ostensibly departed significantly from common sense are worth exploring. DeWeerd [9] found that when predicting the future, readers tended to reject all predictions which didn't match their own and yet were willing to accept at the hand of history what they would instantly reject if predicted in advance by someone else (examples of "unthinkable" outcomes may include the result of the Brexit referendum, the election of Donald Trump, or the attack on Pearl Harbor). Edge-case scenarios are a key virtue of role-plays since they throw up black-swan events – unprecedented and low probability, but high impact events which are invisible to canonical forecasting techniques.

# 3 Methods

The first task when planning a game is to identify the objective (i.e. training, educational, exploring futures or a mix of these). The project to date has explored two different methodological approaches to the game, with slightly different objectives: unstructured and semi-structured games. In both cases, the aim of the game was to allow players to experience from a first-person perspective significant changes that could be brought by the development and deployment of AI technologies, and to have players face strategically challenging decisions from a position of power, be it governmental, corporate, military, societal or other.

## 3.1 Unstructured role-play

For the first two years of the project, we explored various forms of unstructured role-play. Players were drawn from a variety of different backgrounds with varying AI and geopolitical expertise. The players choose their own roles and took unconstrained actions, limited only by the time available and their own creativity. Resolution of actions thus depended heavily on the "world model" of the game's facilitator, which we aimed to address through regular engagement with the AI research literature and the AI strategy research field. The overall aim of these role-plays was to trial out the methodology, and to give players a space in which to explore strategic questions that they found interesting. Whenever it was possible, early discussions with the players or a coordinator helped to determine the initial frame for the scenario and roles adopted by the players. If this was not possible, these would be chosen by the facilitator or in collaboration with the players at the beginning of the game. We experimented with different ways to brief the players, including the facilitator meeting with each player ahead of the session (for 15-30 minutes each) to discuss the strategic questions they are interested in and to flesh out the role for that player. In an alternative approach, the facilitator would tell players the available roles and provide the general framing of the game, before asking the players to do their own background reading on those roles.

Which organizations and roles will matter for the future of AI is an open research question: as time goes by, the number of identified stakeholders only increases, from governments to corporations, militaries to NGOs, artists, environmentalists, criminals and citizens, AI will shape, and be shaped by, a vast number of stakeholders. Nonetheless, the power distribution amongst these actors is not equal, and in varying the organizations and roles in the unstructured role-play we were able to explore different narratives that emerge when different groups and types of power are at play. Over 30 games we trialed many different roles. These broadly clustered around:

- Government organizations and roles in different countries, including the executive, the military, diplomatic services, health, and government R&D functions.
- Corporate actors in different geographies and sectors, including tech (based on business models we see around the world, e.g. advertising, e-commerce or tech infrastructure), robotics, weapons, health.
- Non-governmental actors, including academia, independent research organizations, advocacy groups and think tanks.
- Supra-national actors, e.g. the EU or the UN.

During the game, the participants would "assume" their chosen/assigned role and take actions according to their model of that actor. Participants are instructed to deliberately try to bring about a world that is in line with the goals of the role they are playing, but they should be mindful of real-world constraints. This is because the more closely the simulation captures essential behavioral features of the real system, the more likely insights are to be relevant. On the other hand, a key strength of role-plays is in exploring edge-case scenarios. Thus, as a general rule of thumb, actions should be realistic but not overly conservative. For example, one in ten actions should be an action that the participant assigns 10% likelihood of being taken. Participants are encouraged to "flesh out" the world model by adding details that complement the details provided by the facilitator.

During previous games, participants tried out a wide range of actions, including:

- Launching AI R&D projects
- Deploying AI in various sectors (health, finance, education, defense, cyber)
- Regulating AI technologies, AI products, data or AI talent
- Taxing or breaking up tech companies
- Lobbying governments
- Incentivizing AI talent to switch country, company or project
- Negotiating collaborations on data, funding, talent, hardware, software, etc.
- Gaining access to information and IP through acquisitions, collaborations and espionage
- Interfering with technology-mediated public discourse, or defending against such interventions
- Interfering with digital systems, or defending against such interferences
- Investing in, scaling up, or interfering with AI-specific hardware

Games typically involve 4-8 turns, with each turn representing one or two years, thus covering a period of 4-16 years, with the starting point set in the foreseeable future (as near as 2020 or as far as 2028). The game proceeds by alternating between negotiation phases, private action phases, public actions phases, and "world updates" from the facilitator. Outcomes of actions are determined by the facilitator, sometimes assisted by a random number generator, and relayed either to teams directly or in the "world updates" at the beginning of the next turn. Figure 1 depicts the final state of the world for a recent game held in Cambridge.

> **2035 State of the World:**
> The unemployment rate is now 17.1%. However, people are content for the most part because of social services such as Universal Basic Income. Multiple sources have confirmed to the New York Times that AGI has been developed. Sources range in assessing the system's safety, from 25% to 75% confidence. China is preparing to establish a small city on Mars. The first

Figure 1: Final state of the world from a recent unstructured game. The game involved three players playing the roles of presidents for the USA, PRC and EU. Technological progress and impacts of players' actions were determined by the facilitator.

Immediately at the end of the scenario role-play there is a debrief where any hidden state of the world is revealed, and participants evaluate how well their assigned role performed relative to their goal. Participants are asked to reflect on the final state of the world, on actions or developments that they found surprising, and to record any disagreements with how actions were chosen or resolved. This phase of gleaning participants' insights is crucial for challenging the facilitator's own subjective biases or knowledge gaps whilst bolstering the participants' internalization of what they have learned.

### 3.2 Semi-structured role-play

Based on our experience with unstructured role-play, we started developing a semi-structured game that involves a pre-defined game system for AI R&D, and a pre-defined set of organizations and characters, to create an experience that explores several questions we have found to be particularly pertinent, including:

- How do the geopolitical relations between the two leading AI powers, US and China, affect the kinds of technologies developed and the way they are deployed?
- How do state-corporate relationships play out as AI technologies become more advanced and allow for a greater generation and consolidation of wealth?
- How do states make decisions about the deployment of AI capabilities in strategic domains such as military and social control?
- How do corporations balance the pursuit of risky basic research that could lead to transformative capabilities with nearer term, profit generating products and services?
- How, and to what extent, are states willing to "pick up the tab" for negative externalities created by their corporate champions?
- What factors contribute to a decision to make a research project private/secret, or conversely to publish results and share capabilities? How do actors respond when they suspect other actors of pursuing advanced research in secret?
- How do decisions and impacts relating to near-term AI technologies affect the landscape in which decisions are made about mid-term AI technologies, and similarly how do those affect the setting for decisions about long-term AI?

To explore these questions, we designed an eight-player game with the following roles:

- PRC government: president and minister of national defense
- US government: president and secretary of defense
- Tencent: CEO and head of AI lab
- Alphabet: CEO and head of AI lab

We also created organization sheets and character sheets with key information to familiarize players with their roles.

As in the unstructured game, players were asked to spend each turn negotiating with each other before deciding upon a set of private and public actions. However, we made AI R&D actions structured around a technology tree (Figure 2) and a set of products and capabilities (Figure 3).

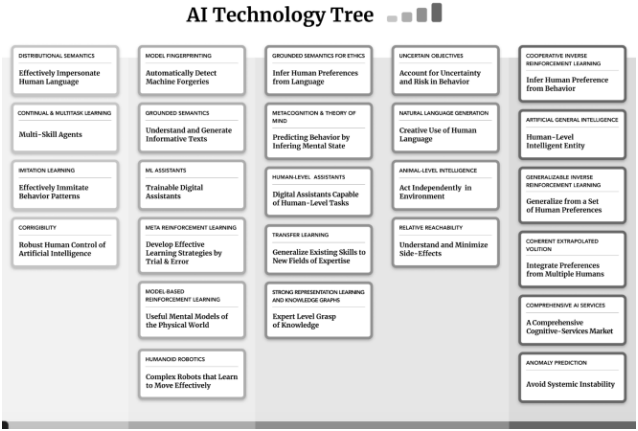

Figure 2: AI Technology Tree used in the game. Each box represents a technology that can be pursued by the players, with associated products and capabilities. Technologies advance from left to right, with progress on earlier technologies required before later technologies are "unlocked".

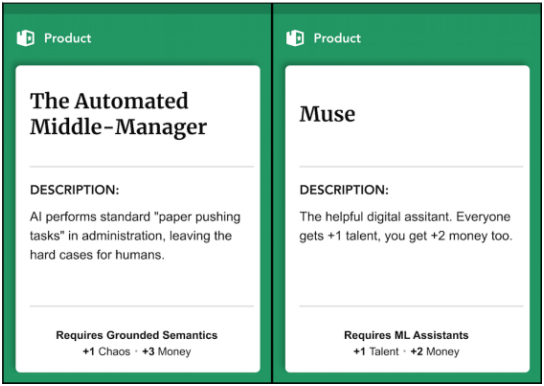

Figure 3: An example of AI products a player can develop in the game. The required technology and the effect on game parameters are listed at the bottom of the card.

Each turn, the different organizations have a limited amount of AI "talent" they can allocate towards pursuing technological advancement (research) or products and capabilities (development). This can be pursued in public or in private. R&D success is determined by dice rolls, with one dice per "talent" allocated to a project. The players' action can affect the distribution of "talent" in the "world". This resource management mechanic was added to explore conditions of competition and scarcity in the input to AI R&D. This is one such factor that generates an adversarial tension which drives players to seek more creative solutions.

Another mechanic we added to the game was "world chaos", a catch-all parameter for global instability, whereby we capture various negative externalities of AI, including the potential of AI "hype" to distract states from addressing issues such as inequality or climate change.

## 4 Preliminary Results

The unstructured game format has allowed the exploration of many different scenarios, but consequently it is difficult to identify patterns in the scenarios played. Overall, participant satisfaction appears high, with most players indicating their interest in playing again, and many interested in reading more about different organizations, roles, actions or strategic analyses. The following range of topics have come up in at least one of the scenarios that emerged from unstructured role-play (this list is not exhaustive):

- International agreements on AI
- Cross-industry agreements on AI
- Cross-industry collaborations on AI safety
- Secret AI R&D
- Espionage
- AI talent poaching
- Hardware supply chain
- AI and the environment
- AI and health
- AI and the economy
- AI and unemployment
- Inequality and unrest
- AI wealth redistribution
- AI and religion
- AI and new social movements
- AI in games and entertainment
- Social robots
- Military AI and autonomous weapon systems
- AI-based cyber capabilities
- AI used for manipulation and disinformation
- State and corporate surveillance with AI

It would often be the case that five or more of the above topics would be touched upon in a single game.

The semi-structured game is more constrained, and therefore provides a better opportunity to detect patterns of gameplay and decision-making. While the number of games played with the new system to date is small, and the game system is still under active development, a few patterns are beginning to emerge.

First, and perhaps not surprisingly, the familiarity of players with the roles assigned to them varies greatly – while many have at least some idea of what the President of the United States cares about or what they can do, that is not the case for tech CEOs or heads of AI labs; while a good facilitator can use the game to provide some basic information about the role and some of the decisions they face (as they relate to AI), the game is more useful for all participants when the players start the game with more background knowledge, both about AI and about the contexts in which decisions about AI are made.

Second, the rapid growth of options that comes with the proliferation of available AI capabilities can be overwhelming to players, who respond to the large number of available R&D avenues by either choosing randomly, picking quickly a "good enough" target, or aiming for a specific long-term technology. While this is to some extent an artefact of the game mechanics, we think it may capture a real issue, especially if AI progress lends itself to faster AI R&D and if progress remains stochastic and uncertain.

Third, cooperation between actors on AI safety and (some) restriction on destabilizing uses of AI seem to both be robustly beneficial, and often necessary, for all players to achieve a good outcome. Such agreements are easier to establish if players are biased towards pro-social strategies, or when cooperation begins early in the game and is maintained and expanded, or when players recognize a common threat (e.g. an unsafe or destabilizing technology nearing completion or after a major accident). Competitive behavior, including adversarial actions (espionage, sabotage, poaching, tariffs), can sometimes be excused when a common threat emerges and the need for cooperation becomes apparent.

Fourth, in almost all games players will continue pursuing ever-more-advanced technologies and capabilities, with all the resources they can afford to spend on R&D, even if the potential value of these advances is unclear, and even when earlier advances have proven to be destabilizing or risky.

Fifth, while the game has a built-in system for AI R&D, some players are happy to innovate "around" this system, proposing new products or establishing new organizations during the game.

Finally, different sets of possible coalitions and alliances are explored and can prove stable, including between government and corporate entities within a state, between governments, between organizations, and between sub-roles within organizations (e.g. between heads of AI labs).

## 5 Discussion and Conclusions

The AI role-play game described here presents an innovative methodology for studying and teaching the impacts of AI. It can educate, raise awareness and help illuminate the positive and negative effects of certain developments, strategic decisions, and inter-stakeholder interactions thereby potentially improving the outcomes of deploying current and future AI technologies. The versatility of the technique, including its ability to be adapted to different problems throughout organizations to address the near-, mid- and long-term impacts of AI, is well-suited to the large variety of challenges encountered in the development and deployment of AI technologies across all aspects of human life. Many of the other benefits of role-play games discussed in the literature review are also applicable.

The games we have conducted to date, and especially the semi-structured games, focused on a set of questions and insights that we find particularly salient. These include the interaction between near-, mid- and long-term risks, challenges of cooperation and competition between powerful actors across states of public/private lines, trade-offs between public and private research and between research and product development, and the experience of making potentially transformative decisions from a first-person (powerful) perspective under conditions of uncertainty and time pressure. The feedback from players to date suggests that these aspects of the game indeed appear salient to players, and lead to greater engagement with these questions.

Based on players' feedback, and from design constraints of the game itself, we also note several shortcomings and limitations we have not yet been able to address. For example, in our focus on strategic decision making, we have focused on role-play from the perspective of powerful actors, rather than actors with marginalized standpoints, or more detached viewpoints that highlight systemic and emergent drivers; the literature suggests that role-plays and games can be adapted for these perspectives, and have been powerful in building empathy or systems thinking when done well, and this is a potential direction for expansion.

Another limitation is the sped-up timeline, that imagines dramatic, and occasionally transformative, technological shifts within an almost-foreseeable political horizon. This reflects our limited ability to conduct a realistic role-play that takes place in the political reality of 2050 or 2100, but it does mean we adopt an assumption of fairly rapid technology progress, which is in no way certain – experts significantly disagree about timelines to transformative AI capabilities and impacts [12, 13]. This design decision was made in part to play to the strength of role-plays, which shine in exploring large and uncertain effects, partly to create a more engaging and dramatic setting, and partly because if transformative impacts happen in a matter of decades, it becomes of paramount importance to prepare decision-makers to handle the decisions they will need to face. We also stress the communication of this bias in every game debrief, but we see how this sped-up timeline may be less appealing to some, and are exploring versions of the game that focus on less dramatic changes while exploring in greater nuance short- and medium-term impacts of technologies that are already making real research progress.

Moving forward, we will continue to develop the game to increase realism, reflect new strategic research questions as they emerge, and cover more scenarios of interest to a broader set of stakeholders. We aim to expand the context of the game, through different unstructured and semi-structured scenarios and mechanics, to cover an increasing range of actors, actions, technologies and impacts of AI. This necessitates staying in constant contact and mutual feedback with the developing fields of AI futures and AI strategy as the impacts of AI become more apparent in every corner of life. In addition, we hope to continue to develop the educational potential of the game into something that can be used to raise awareness and train a variety of audiences and publics, including younger members of the public, through the development of a digital platform and a digital version of the game. Readers interested in exploring the game are invited to register their interest,[1] and those who wish to join the development of the game are encouraged to contact the authors.

By allowing stakeholders to explore alternative AI futures, prepare for surprises and uncertainty, and learn from "simulated" mistakes, we hope that the methodology we propose will help win the race between the capability of the technologies we develop and our wisdom in deploying and governing them. For these reasons we see the method's impact on the use of AI for social good as being far reaching and we invite members of the community to join us in exploring the future of AI through role-play.

## ACKNOWLEDGMENTS

This project is supported by a grant from the Centre for Effective Altruism Long Term Future Fund, awarded to Shahar Avin.

## REFERENCES

[1] Anderson, R. H., and Hearn, A. C. 1997. An Exploration of Cyberspace Security R&D Investment Strategies for DARPA. Santa Monica, CA: RAND Corporation.
[2] Armstrong, J. S. 2001 Role Playing: A Method to Forecast Decisions. In *Principles of Forecasting*: 15–30. Boston, Mass: Springer.
[3] Avin, S. 2019. Exploring Artificial Intelligence Futures. *Journal of AI Humanities*, 35812.
[4] Babcock, L., Loewenstein, G., Issacharoff, S., and Camerer, C. 1995. Biased Judgments of Fairness in Bargaining. *The American Economic Review* 85(5): 1337–1343.
[5] Cave, S., and O'hEigeartaigh, S. 2019. Bridging Near- and Long- term Concerns About AI. *Nature Machine Intelligence* (1): 5-6.
[6] Cook, M.P., Gremo, M. and Morgan, R., 2017. We're just playing: The influence of a modified tabletop role-playing game on ELA students' in-class reading. *Simulation & Gaming*, 48(2), pp.199-218.
[7] Cyert, R. M., March, J. G., and Starbuck, W. H. 1961. Two Experiments on Bias and Conflict in Organizational Estimation. *Management Science* 7(3): 254–264.
[8] Dafoe, A. 2018. AI governance: A research agenda. *Governance of AI Program, Future of Humanity Institute, University of Oxford: Oxford, UK.*
[9] DeWeerd, H. A. 1974. A Contextual Approach to Scenario Construction. *Simulation & Games* 5(4): 403–414.
[10] Duke, R. D., and Geurts, J. L. A. 2004. *Policy Games for Strategic Management: Pathways Into the Unknown.* Oisterwijk, The Netherlands: Dutch University Press.

---

[1] To register your interest in participation, please fill out the form at https://sites.google.com/view/intelligence-rising/home.

[11] Green, K. C., Graefe, A., and Armstrong, J. S. 2011. Forecasting Principles. In *International Encyclopedia of Statistical Science*. Ed. M. Lovric. Boston, Mass: Springer.

[12] Grace, K., Salvatier, J., Dafoe, A., Zhang, B. and Evans, O. 2018. When will AI exceed human performance? Evidence from AI experts. *Journal of Artificial Intelligence Research*, *62*, pp.729-754.

[13] Gruetzemacher, R., and Paradice, D. 2019. Alternative Techniques for Mapping Paths to HLAI. *arXiv preprint arXiv:1905.00614*.

[14] Gruetzemacher, R., and Whittlestone, J. 2019. Defining and Unpacking Transformative AI. (Working paper)

[15] Hamel, G., and Prahalad, C. K. 1996. *Competing for the Future*. Cambridge, Mass: Harvard Business Review Press.

[16] Kanner, M.D., 2007. War and peace: Simulating security decision making in the classroom. *PS: Political Science & Politics*, *40*(4), pp.795-800.

[17] Lipinski, H., Amara, R., and Spangler, K. 1978. Communication Needs in Computer Modeling. In *Proceedings of the 10th conference on winter simulation* (1):219–228.

[18] Nestel, D. and Tierney, T., 2007. Role-play for medical students learning about communication: guidelines for maximising benefits. *BMC medical education*, *7*(1), p.3.

[19] Parson, E. A. 1996. What Can you Learn from a Game? In *Wise Choices: Games, Decisions, and Negotiations*. Eds. R. Zeckhauser, R. Kenney & J. Sebenius. Cambridge, Mass: Harvard Business School Press.

[20] SIGNAL. 2019. Signal video game. Retrieved from https://www.signalvideogame.com/ on 6th September 2019.

[21] Sogunro, O.A., 2004. Efficacy of role-playing pedagogy in training leaders: some reflections. *Journal of management development*.

[22] Starkey, B. A., and Blake, E. L. 2001. Simulation in International Relations Education. *Simulation & Gaming* 32(4): 537–551.

[23] Wheeler, S.M., 2006. Role-playing games and simulations for international issues courses. *Journal of Political Science Education*, *2*(3), pp.331-347.

[24] Wilkinson, S. 1887. *Essays on the War Game*. Manchester, United Kingdom: Manchester Tactical Society.